\documentclass{PoS}

\title{Hamiltonian Dyson--Schwinger Equations of QCD}

\ShortTitle{Hamiltonian Dyson--Schwinger Equations of QCD}

\author{\speaker{Davide Campagnari} and Hugo Reinhardt\\
        Institute for Theoretical Physics, University of T\"ubingen\\
	Auf der Morgenstelle 14, 72076 T\"ubingen, Germany\\
        E-mail: \email{d.campagnari@uni-tuebingen.de}}

\abstract{The general method for treating non-Gaussian wave functionals in the Hamiltonian
formulation of a quantum field theory, which was previously developed and applied to
Yang--Mills theory in Coulomb gauge, is generalized to full QCD. The Hamiltonian
Dyson--Schwinger equations as well as the quark and gluon gap equations are derived and analysed in the
IR and UV momentum regime. The back-reaction of the quarks on the gluon sector is investigated.}

\FullConference{Xth Quark Confinement and the Hadron Spectrum,\\
		October 8--12, 2012\\
		TUM Campus Garching, Munich, Germany}

\usepackage{amsmath}
\DeclareMathOperator{\Det}{Det}	
\newcommand*{\valpha}{{\vec{\alpha}}}
\newcommand*{\vx}{{\vec{x}}}
\newcommand*{\vy}{{\vec{y}}}
\newcommand*{\vz}{{\vec{z}}}
\newcommand*{\vp}{{\vec{p}}}
\newcommand*{\vq}{{\vec{q}}}

\newcommand*{\vA}{{\vec{A}}}
\newcommand*{\uvp}{{\Hat{p}}}
\newcommand*{\uvq}{{\Hat{q}}}
\newcommand*{\I}{\ensuremath{\mathrm{i}}}%
\newcommand*{\be}{\begin{equation}}
\newcommand*{\ee}{\end{equation}}
\newcommand*{\braket}[2]{ \langle #1 \vert #2 \rangle }
\newcommand*{\bra}[1]{\langle #1 \rvert \mkern2mu}
\newcommand*{\ket}[1]{\mkern2mu \lvert #1 \rangle}
\newcommand*{\tsqrt}[1]{\sqrt{\smash[b]{#1}}}
\makeatletter
\def\vev{\@ifstar\@vev\@@vev}
\newcommand*{\@vev}[1]{\langle #1 \rangle}
\newcommand*{\@@vev}[1]{\left< #1 \right>}
\def\Eqref{\@ifstar\@Eqref\@@Eqref}
\newcommand*{\@Eqref}[1]{Eq.\ \eqref{#1}}
\newcommand*{\@@Eqref}[1]{Eq.~\eqref{#1}}
\makeatother
\newcommand*{\calD}{\ensuremath{\mathcal{D}}}
\newcommand*{\calE}{\ensuremath{\mathcal{E}}}
\newcommand*{\Nc}{\ensuremath{N_\mathrm{c}}}
\newcommand*{\calO}{\ensuremath{\mathcal{O}}}
\newcommand*{\e}{\ensuremath{\mathrm{e}}}

\newcommand*{\eqcolon}{=\mathrel{\mathop:}}
\renewcommand*{\d}[1][]{\mathop{\mathrm{d}^{#1}}\mkern-4mu} 
\newcommand*{\abs}[1]{\ensuremath{\lvert#1\rvert}}
\newcommand*{\dfr}[2][]{{\ifx&#1&\frac{\mathrm{d}#2}{2\pi}\else\frac{\mathrm{d}^{#1}#2}{(2\pi)^{#1}}\fi}}

\begin{document}

\section{Introduction}

Coulomb gauge Yang--Mills theory has attracted a considerable amount of attention over
the last few years. In the continuum, both the Hamiltonian \cite{Schutte:1985sd,Szczepaniak:2001rg,Feuchter:2004mk,Epple:2006hv}
and the Lagrangian \cite{Watson:2006yq,Watson:2007vc,Reinhardt:2008pr,Watson:2010cn} approach have
been investigated. In the Hamiltonian approach, in particular, the use of variational
methods with Gaussian wave functionals has led to the so-called gap equation for the
inverse equal-time gluon propagator.
The analytical and numerical solutions show an inverse gluon propagator which in the UV behaves like the photon
energy but diverges in the IR, signalling confinement. The obtained propagator also compares
favourably with the available lattice data \cite{Burgio:2008jr}; deviations in the
mid-momentum regime (and minor ones in the UV), which can be attributed
to the gluon loop escaping the Gaussian wave functionals, can be taken into consideration
by going beyond the purely Gaussian form. To treat these
non-Gaussian functionals the authors developed in Ref.~\cite{Campagnari:2010wc} a method
relying on Dyson--Schwinger Equations (DSEs) by exploiting the formal similarity between
vacuum expectation values in the Hamiltonian formalism and correlation functions in
Euclidean quantum field theory. In this talk we report about the implementation of these
techniques in the quark sector of the theory, in order to investigate issues like the
spontaneous breaking of chiral symmetry.

The topic of chiral symmetry breaking in Coulomb gauge has been studied for example in
Ref.~\cite{Adler:1984ri}. While it has been shown that an infrared enhanced
potential can account for chiral symmetry breaking, the calculated physical quantities,
such as the dynamical mass and chiral condensate, turn out to be far too small.
It has been recently shown \cite{Pak:2011wu} that using a wave functional which includes
the coupling of the quarks to the transverse gluon field improves the results towards the phenomenological findings.
We apply here the techniques developed in Ref.~\cite{Campagnari:2010wc}
to the wave functional proposed in Ref.~\cite{Pak:2011wu}. While we do not expect considerably
different results from Ref.~\cite{Pak:2011wu} for physical quantities, the approach presented here has a broader range of
applications, and could easily be applied to more complicated wave functionals.


\section{Hamiltonian Dyson--Schwinger Equations}

The derivation of the Hamiltonian DSEs in pure Yang--Mills theory has been presented in
Ref.~\cite{Campagnari:2010wc}. We summarize here briefly the essential steps leading from the
choice of the wave functional to the pertinent DSEs.

In the Schr\"odinger picture of the Yang--Mills sector we represent the gauge field and
the canonical momentum operators in the basis $|A\rangle$ of the eigenstates of the
field operator by
\be\label{schroe1}
\bra{A} \hat{A} \ket{\phi} = A \, \varPhi[A] , \qquad
\bra{A} \hat{\Pi} \ket{\phi} = \frac{\delta}{\I \, \delta A} \: \varPhi[A] ,
\ee
where $\varPhi[A] \equiv \braket{A}{\varPhi}$ is a physical state. The vacuum expectation
value (VEV) of an operator $K$ is therefore given by
\be\label{mad1}
\vev{K[A,\Pi]} = \int \calD A \: J_A \: \varPhi^*[A] \: K[A,\tfrac{\delta}{\I\delta A}] \: \varPhi[A].
\ee
In \Eqref{mad1} the functional integration runs over transverse field configurations
satisfying the Cou\-lomb gauge condition, $\partial_i A_i^a=0$, and is restricted to the
first Gribov region;  $J_A=\Det(G_A^{-1})$ is the Faddeev--Popov determinant which arises
from the gauge fixing, and
\be\label{cx2}
\bigl[G_A^{ab}(\vx,\vy)\bigr]^{-1} = \bigl( - \delta^{ab} \partial^2 - g f^{acb} A_i^c(\vx) \partial_i \bigr) \delta(\vx-\vy)
\ee
is the inverse Faddeev--Popov operator, with $g$ being the coupling constant and $f^{abc}$
being the structure constants of the $\mathfrak{su}(\Nc)$ algebra.

In a similar manner, a state $\lvert\varPhi\rangle$ in the fermionic Fock space possesses
a ``coordinate'' representation
\be\label{dir3}
\braket{\xi}{\Phi} \equiv \varPhi(\xi^\dag,\xi) ,
\ee
where $\xi$, $\xi^\dag$ are complex Grassmann fields.
The VEV of a fermionic operator in this state is given by
\be\label{dir4}
\vev{ \calO[\hat\psi,\hat\psi^\dag] } = \int \calD\xi^\dag \, \calD\xi \:
\e^{-\xi^\dag(\Lambda_+-\Lambda_-)\xi} \: \varPhi^*(\xi,\xi^\dag) \:
\calO \bigl[ \hat{\psi},\hat{\psi}^\dag \bigr] \:
\varPhi(\xi^\dag,\xi) ,
\ee
where the $\Lambda_\pm$ are the projectors onto states of positive and negative energy
of the free Dirac theory. The Dirac field operators act on functionals according to
\be\label{vev3}\begin{split}
\hat{\psi} \varPhi(\xi^\dag,\xi) &= \left( \Lambda_- \xi + \Lambda_+\frac{\delta}{\delta \xi^\dag}\right) \varPhi(\xi^\dag,\xi) , \\
\hat{\psi}^\dag  \varPhi(\xi^\dag,\xi) &= \left( \Lambda_+\xi^\dag + \Lambda_-\frac{\delta}{\delta \xi} \right) \varPhi(\xi^\dag,\xi) .
\end{split}
\ee
We have put explicitly a hat over the fermion operators $\hat\psi^\dag$ in
\Eqref{vev3} to distinguish them
from the Grassmann (classical) fields $\xi$, $\xi^\dag$ used in the ``coordinate'' representation.
The exponential factor occurring in the fermionic functional integration in \Eqref{dir4}
arises from the completeness relation for fermionic coherent states, see e.g.~Ref.~\cite{Berezin:1966nc}.

The vacuum state of QCD can be assumed to be of the form
\be\label{qcd0}
\varPsi[A,\xi,\xi^\dag] \eqcolon \exp\left\{ -\frac12 \, S_A[A] - S_f[\xi,\xi^\dag,A] \right\} ,
\ee
where $S_A$ is a functional of the gauge field only, while $S_f$ contains
the fermion and fermion-gluon interaction parts. Dyson--Schwinger equations are derived
by starting from the identity \cite{Campagnari:2010wc}
\be\label{cx3}
0 = \int \calD A \, \calD\xi^\dag \, \calD\xi \: \frac{\delta}{\delta\phi}
\left\{ J_A \, \e^{-\xi^\dag(\Lambda_+-\Lambda_-)\xi} \:
\varPsi^*[A,\xi,\xi^\dag] \, K[A,\xi,\xi^\dag] \, \varPsi[A,\xi,\xi^\dag] \right\}
\ee
with $\phi\in\{A,\xi,\xi^\dag\}$.

To proceed further, we need an explicit ansatz for the vacuum wave functional \Eqref{qcd0}. Since
we are interested here mainly in the quark sector, we choose the simple Gaussian form
\be\label{cx4}
S_A = \int \d[3]x \, \d[3]y \: A^a_i(\vx) \, \omega(\vx,\vy) \, A_i^a(\vy)
\ee
for the Yang--Mills part.
It should be stressed, however, that this is in no way necessary, and non-Gaussian generalization
can be taken as well \cite{Campagnari:2010wc}; we restrict ourselves to the form \Eqref{cx4} only 
for pedagogical reasons.

For the quark wave functional we choose the ansatz \cite{Pak:2011wu}
\be\label{cx5}
S_f = \int \d[3]x \, \d[3]y \: \xi_+^{m\dag}(\vx) \, K_A^{mn}(\vx,\vy) \, \xi_-^n(\vy),
\qquad \xi_{\pm}=\Lambda_{\pm}\xi.
\ee
The kernel $K_A$ contains both
the purely fermionic part and the coupling of the quarks to the transverse gluons,
\be\label{cx6}
K_A^{mn}(\vx,\vy) = \delta^{mn} \beta \, s(\vx,\vy) + g \int \d[3]z \: V(\vx,\vy;\vz) \, \valpha\cdot\vA^a(\vz) \: t_a^{mn} .
\ee
Here, $\beta$ and $\valpha$ are the usual Dirac matrices, $t_a$ are the generators of
the group in the fundamental representation (or, in general, in the representation to which
the fermion fields belong), and the functions $s$ and $V$ are the variational kernels.

\begin{figure}
\centering
\parbox{.45\linewidth}{ \includegraphics[width=\linewidth]{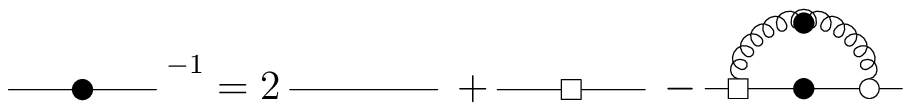} \\ \vfill }
\null\hfill
\parbox{.45\linewidth}{ \includegraphics[width=\linewidth]{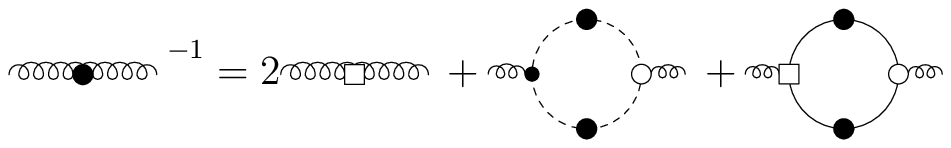} }
\caption{\label{fig:DSEs}Hamiltonian DSEs for the quark (left) and gluon propagator (right).
Filled dots stand for dressed propagators and empty circles for proper functions. The small empty
boxes represent the variational kernels. Curly lines represent gluons, dashed lines ghosts,
and straight lines fermions.}
\end{figure}

With these choices for the wave functional the resulting DSEs are represented diagrammatically
in Fig.~\ref{fig:DSEs}.


\section{Variational Approach to QCD}

The Hamilton operator of QCD in Coulomb gauge \cite{Christ:1980ku}, resulting from the
resolution of Gauss's law in the canonically quantized theory, can be written as
\begin{multline}\label{cx1}
H_{\mathrm{QCD}} =
- \frac12 \int \d[3]x \: J_A^{-1} \frac{\delta}{\delta A_i^a(\vx)} J_A \frac{\delta}{\delta A_i^a(\vx)}
+ \frac14 \int \d[3]x \: F_{ij}^a(\vx) \, F_{ij}^a(\vx) \\
+\int \d[3]x \: \psi^{m\dag}(\vx) \bigl[-\I \alpha_i \partial_i + \beta \, m\bigr] \psi^m(\vx)
- g \int \d[3]x \: \psi^{m\dag}(\vx) \alpha_i A_i^a(\vx) \, t_a^{mn} \psi^n(\vx) \\
+ \int \d[3]x \d[3]y \: J_A^{-1} \rho^a(\vx) J_A \, F_A^{ab}(\vx,\vy) \, \rho^b(\vy) .
\end{multline}
In \Eqref{cx1}
\be\label{fst}
F_{ij}^a(\vx) = \partial_i A_j^a(\vx) - \partial_j A_i^a(\vx) + g f^{abc} A_i^b(\vx) \, A_j^c(\vx)
\ee
is the spatial part of the field strength tensor,
\be\label{coulkernel}
F_A^{ab}(\vx,\vy) = \int \d[3]z \: G_A^{ac}(\vx,\vz) \, (-\partial^2_z) \, G_A^{cb}(\vz,\vy)
\ee
is the so-called Coulomb kernel, which arises from the longitudinal component of the
electric field, and
\be\label{charge}
\rho^a(\vx) = \psi^{m\dag}(\vx)t_a^{mn}\psi^n(\vx) + f^{abc} A_i^b(\vx) \frac{\delta}{\I \, \delta A_i^c(\vx)}
\ee
is the total colour charge density.

The energy density can be evaluated as expectation value of the Hamiltonian \Eqref{cx1}.
The Hamiltonian DSEs in the general form \Eqref{cx3} enter this calculation twice: first
to express the VEVs of the operators $\hat\psi$, $\hat\psi^\dag$ in terms of the Grassmann
fields $\xi$, $\xi^\dag$, and second to turn the resulting expressions, which involve
propagators and vertex functions, in a functional of the variational kernels. In this
calculation we have to introduce approximations, and we evaluate the energy density up to
two loops (see Ref.~\cite{Campagnari:2010wc} for more details).


\section{Gap Equations}

The variation of the energy density with respect to $V$ fixes the vector kernel
\be\label{loqcd5}
V(\vp,\vq) = - \frac{1+s(\vp) \, s(\vq)}{\Omega(\vp+\vq) + \abs{\vp} \frac{1-s^2(\vp)+2 s(\vp) s(\vq)}{1+s^2(\vp)} + \abs{\vq} \frac{1-s^2(\vq)+2 s(\vp) s(\vq)}{1+s^2(\vq)} } ,
\ee
where $\Omega=\tfrac12 \vev*{AA}^{-1}$ is the inverse gluon propagator.
Replacing the kernels occurring on the right-hand side by their tree-level forms, i.e.~$\Omega(\vp)=\abs{\vp}$
and $s(\vp)=0$, we obtain
\be\label{cx7}
V_0(\vp,\vq) = - \frac{1}{\abs{\vp+\vq} + \abs{\vp} + \abs{\vq} },
\ee
which is exactly the leading-order perturbative expression for the quark-gluon vertex
for massless fermions \cite{DissCampagnari}.

Equation \eqref{loqcd5} can be inserted back into the expression for the energy density, and taking
functional derivatives with respect to $\omega(\vp)$ and $s(\vp)$ we obtain the gluon
and quark gap equations. As a first approximation, we will ignore the $s$-dependence of
the denominator of \Eqref{loqcd5}. Keeping the whole denominator structure would give rise to more
complicated expressions which, however, have the same IR and UV asymptotic behaviour.

The approximated gap equation for the scalar quark kernel $s$ reads
\be\label{gap2p}
\begin{split}
\abs{\vp} \, s(\vp) 
={}& \frac{g^2 C_F}{2} \int \dfr[3]{q} \; F(\vp-\vq) \:
\frac{s(\vq) \bigl[1-s^2(\vp)\bigr] - \uvp\cdot\uvq \, s(\vp) \bigl[1-s^2(\vq)\bigr]}{1+s^2(\vq)} \\
&+ \frac{g^2 C_F}{2} \int \dfr[3]{q} \:
\frac{X(\vp,\vq)}{\Omega(\vp+\vq)}
\frac{\bigl[1+s(\vp)\,s(\vq)\bigr] \bigl[s(\vq)-s(\vp)\bigr]}{ \bigl[ 1+s^2(\vq) \bigr] \, \bigl[\Omega(\vp+\vq) + \abs{\vq} + \abs{\vp} \bigr]} ,
\end{split}
\ee
where $F=\vev*{F_A}$ is the Coulomb propagator and $X$ is a tensor structure arising
from the traces in Dirac space.
The first term on the right-hand side of \Eqref{gap2p} was obtained by
Adler and Davis \cite{Adler:1984ri}; the second term is due to the coupling of the quarks to
the transverse gluons. As mentioned in the Introduction, in Ref.~\cite{Pak:2011wu} the
coupling to transverse gluons was also taken into account. The authors of Ref.~\cite{Pak:2011wu}
used the same wave functional \Eqref{cx5} as in this work but did a quenched calculation.
As a consequence the effect of the Yang--Mills kinetic energy operator on the quarks ws neglected.

Trading the scalar kernel $s$ in favour of the mass function $M$, \Eqref{gap2p} takes
the following simple form
\be\label{gap2a}
\begin{split}
\abs{\vp} M(\vp) ={}& g^2 \frac{C_F}{2} \int \dfr[3]{q} \; F(\vp-\vq)
\frac{ \abs{\vp} M(\vq) - \uvp\cdot\uvq \, \abs{\vq} \, M(\vp)}{\calE(\vq)} \\
&+g^2 \frac{C_F}{2} \int \dfr[3]{q} \: \frac{X(\vp,\vq)}{\Omega(\vp+\vq) \bigl[ \Omega(\vp+\vq) + \abs{\vp} + \abs{\vq} \bigr]}
\frac{\abs{\vp} M(\vq) - \abs{\vq} M(\vp)}{\calE(\vq)} ,
\end{split}
\ee
with
\be
\calE(\vp) = \tsqrt{\vp^2 + M^2(\vp)} .
\ee

The variation of the energy density with respect to the gluon kernel $\omega$ combined with
the gluon propagator DSE yields the gap equation for inverse gluon propagator $\Omega$
\begin{multline}\label{gap1}
\Omega^2(\vp) = \vp^2 + \text{Yang--Mills terms} \\
- g^2 \int \dfr[3]{q}
\frac{ \bigl[ 1+s(\vq) \, s(\vp+\vq) \bigr]^2 }{ \bigl[ 1+s^2(\vq) \bigr] \bigl[1+s^2(\vp+\vq) \bigr] }
\frac{2\Omega(\vp) + \abs{\vq} + \abs{\vp+\vq}}{\bigl[\Omega(\vp) + \abs{\vq} + \abs{\vp+\vq} \bigr]^2}  \\
\times \left[ 1 +\frac{\vp\cdot\vq + (\uvp\cdot\vq)^2}{\abs{\vq} \, \abs{\vp+\vq}} \right] ,
\end{multline}
where the usual Yang--Mills terms (ghost loop, Coulomb interaction, and possibly additional
terms stemming from a non-Gaussian ansatz) can be found in Ref.~\cite{Campagnari:2010wc}.

\begin{figure}
\centering
\includegraphics[width=.5\linewidth]{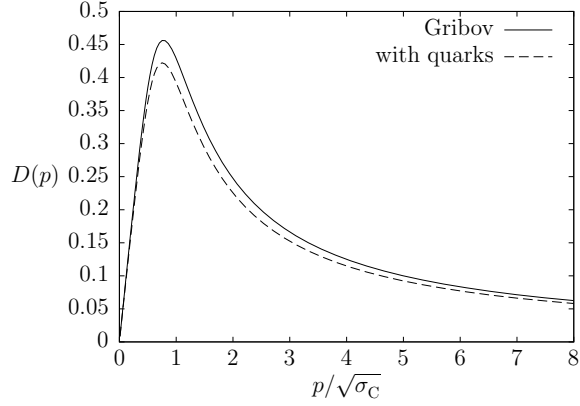}
\caption{\label{fig:gluonprop_quarks}Gluon propagator $D=1/(2\Omega)$ in the presence of
quarks for the gap equation \protect\eqref{gap1} assuming the Gribov form for the Yang--Mills
part. Physical dimensions are set by the Coulomb string tension $\sigma^{}_{\mathrm{C}}$.}
\end{figure}

Equations \eqref{gap2p} and \eqref{gap1} form a coupled system, toghether with the DSE
for the ghost propagator and the Coulomb form factor. The analysis of the full coupled set
of equations is subject of ongoing work. As a first estimate, we investigate the effect of the
quark loop on the gluon propagator: we take the Gribov formula for the
pure Yang--Mills gluon propagator and use for the scalar kernel $s$ a form fitted from
the data in Ref.~\cite{Pak:2011wu}. The result is plotted in Fig.~\ref{fig:gluonprop_quarks}.
As we see, the gluon propagator loses some stregth in the mid-momentum regime. This behaviour
is known from Landau gauge studies, and Coulomb gauge investigations on the lattice
confirm this \cite{privateGB}.


\section{Conclusions}

The general method to treat non-Gaussian wave functionals in the Hamiltonian formulation of a
quantum field theory presented in Ref.~\cite{Campagnari:2010wc} has been applied here to
QCD in Coulomb gauge. We have used the quark wave functional suggested in Ref.~\cite{Pak:2011wu},
which includes the coupling of the quarks to the transverse gluons. The resulting quark gap
equations are similar in both approaches; the numerical solution of
the equations and the comparison to the findings of Ref.~\cite{Pak:2011wu} are the subject
of ongoing work. As a first application we have presented here the effect of the back-coupling
of the quarks on the gluon propagator. In future work we intend to expand the present
approach to QCD at finite temperature \cite{Heffner:2012sx} and baryon density.


\section*{Acknowledgments}
The authors are grateful to M.\ Pak and P.\ Watson for useful discussions. This work was
supported by the DFG under contracts No.\ Re856/9-1 and by BMBF 06TU7199.

\end{document}